\documentclass[10pt,preprint2]{aastex}
\received{2010 June~25}
\revised{2010 September~20}
\accepted{2010 September~20}
\journalid{}{}
\articleid{}{}
\paperid{AJ/360886/ART/176005}
\cpright{}{2010}
\ccc{}
\shorttitle{Radio Observations of HD~80606 Near Planetary Periastron}
\shortauthors{Lazio et al.}

\newcommand{\mjybm}{\mbox{mJy~beam${}^{-1}$}}
\newcommand{\ujybm}{\mbox{$\mu$Jy~beam${}^{-1}$}}
\newcommand{\hdstar}{\protect\objectname{HD~80606}}
\newcommand{\planet}{\protect\objectname{HD~80606b}}

\begin{document}
\title{Radio Observations of HD~80606 Near Planetary Periastron}

\author{T.~Joseph~W.~Lazio\altaffilmark{1,2}}
\affil{Naval Research Laboratory, 4555 Overlook Ave.~\hbox{SW}, 
	 Washington, DC 20375-5351  USA}
\altaffiltext{1}{NASA Lunar Science Institute, NASA Ames Research
  Center, Moffett Field, CA  94035, USA}
\altaffiltext{2}{current address: Jet Propulsion Laboratory, M/S
  138-308, 4800 Oak Grove Dr., Pasadena, CA  91109 USA}

\author{P.~D.~Shankland}
\affil{U.{}S.\ Naval Observatory Flagstaff Station, Flagstaff, AZ 86001 USA}

\author{W.~M.~Farrell\altaffilmark{1}}
\affil{NASA Goddard Space Flight Center, Greenbelt, \hbox{MD} 20771 USA}

\and 

\author{D.~L.~Blank}
\affil{Centre for Astronomy, School of Mathematical and Physical Sciences, James Cook University, Townsville QLD~4811, Australia}

\begin{abstract}
This paper reports Very Large Array observations at~325
and~1425~MHz ($\lambda$90cm and~$\lambda$20cm) during and near the
periastron passage of \planet\ on~HJD~2454424.86 (2007 November~20).
We obtain flux density limits (3$\sigma$) of~1.7~mJy and~48~$\mu$Jy
at~325 and~1425~MHz, respectively, equivalent to planetary luminosity
limits of $2.3 \times 10^{24}$~erg~s${}^{-1}$ and $2.7 \times
10^{23}$~erg~s${}^{-1}$.  Unfortunately, these are several orders of
magnitude above the nominal Jovian value (at~40~MHz) of $2 \times
10^{18}$~erg~s${}^{-1}$.  The motivation for these observations was
that the planetary magnetospheric emission is driven by a stellar
wind-planetary magnetosphere interaction so that the planetary
luminosity would be elevated near periastron.  We estimate that, near
periastron, \planet\ might be as much as 3000 times more luminous than
\objectname{Jupiter}.  Recent transit observations of \planet\ provide
reasonably stringent constraints on the planetary mass and radius,
and, because of the planet's highly eccentric orbit, its rotation
period is likely to be ``pseudo-synchronized'' to its orbital period,
allowing a robust estimate of the former.  Consequently, we are able
to make relatively robust estimates of the emission frequency of the
planetary magnetospheric emission and find it to be around~60--90~MHz.
While this is too low for our reported observations, we compare
\planet\ to other high-eccentricity systems and assess the detection
possibilities for both near-term and more distant future systems.  Of
the known high eccentricity planets, only \planet\ is likely to be
detectable, as the others (\objectname{HD~20782B} and
\objectname{HD~4113}) are both lower mass and longer rotational
periods, which imply weaker magnetic field strengths.  We find that
both the forthcoming ``EVLA low band'' system, which will operate as
low as 65~MHz, and the Low Frequency Array may be able to
improve upon our planetary luminosity limits for \planet, and do so at
a more optimum frequency.  If the low-frequency component of the
Square Kilometre Array (SKA-lo) and a future lunar radio array
are able to approach their thermal noise limits, they should be able
to detect an \planet-like planet, unless the amount by which the
planet's luminosity increases is substantially less than the factor
of~3000 that we estimate; for the SKA-lo, which is to be located in
the southern hemisphere, future planetary surveys will have to find
southern hemisphere equivalents of \planet.
\end{abstract}

\keywords{planetary systems --- planets and satellites: magnetic
  fields --- planets and satellites (\planet) --- radio continuum: planetary systems}

\section{Introduction}\label{sec:intro}

\hdstar\ is a G5 star, at a distance of~58.4~pc, which is distinguished
by being orbited by a planet with one of the highest known orbital eccentricities.  The
planet is a giant planet ($3.94 \pm 0.11\,M_J$) with a 111-day orbital period on an
extremely eccentric orbit ($e=0.93$) \citep{nlm+01,phi+09}.

All of the giant planets in the solar system and the
\objectname{Earth} generate radio emission as a result of an
interaction between the solar wind and the planetary magnetosphere.
Specifically, the solar wind impinging on the planetary magnetosphere
generates currents within the magnetosphere which are then directed
into the planet's magnetic polar regions where an electron cyclotron
maser instability is produced.  Motivated by the radio emission from
solar system planets, there have been a number of searches for
magnetospherically generated radio emission from extrasolar planets,
at frequencies ranging from~25~MHz to~1400~MHz
\citep{yse77,wdb86,bdl00,ztrr01,lf+04,rzr04,gs07,lf07,scgjlb09,ldesg-kz09,l+10}.
Sensitivities (1$\sigma$) have ranged from roughly 1000~mJy to
below~1~mJy, with generally better sensitivities obtained at the higher
frequencies.  Typically, these sensitivities have not been sufficient
to challenge the predicted levels of extrasolar planetary
magnetospheric emission, though some recent multi-epoch 74~MHz
observations of $\tau$~Boo \citep{lf07} do indicate that its planet's
emission is not consistent with the most optimistic predictions,
unless the emission is either at a frequency well below~74~MHz or is
much more strongly beamed than that of solar system planets.

Most recent radio observations of planets have focused on planets in
orbits with relatively small semi-major axes and low eccentricities \citep[but see][]{l+10}.
Part of the motivation for this focus is the stellar wind powering of
the electron cyclotron maser instability.  Close to the host star, the
stellar wind should be more intense with a concomitant increase in the
strength of the planetary radio emission.  A secondary cause is the
result of a selection effect.  Planets close to their host stars
produce the largest radial velocity signature, and the vast majority
of known extrasolar planets have been found with the radial velocity
method.

A concern with this focus on so-called hot Jupiters is that they
are likely to be tidally locked to their host star, with rotation
periods of order a few days.  To the extent that rapid
rotation may be necessary for the generation of a strong planetary
magnetic field, these ``hot Jupiters'' may have weak fields that do
not generate much electron cyclotron maser emission or generate it at
a frequency well below that accessible from the ground.  Absent a
direct radio detection of magnetically generated emission, evidence
for extrasolar planetary magnetic fields is ambiguous.  
Consistent
with the scenario of weak fields, \cite{fsy+10} were unable to find any unambiguous
signal of aurorally generated UV emission from the ``hot Jupiter''
\objectname{HD~209458b}.  From this they derive a limit on the
intrinsic planetary magnetic moment of~0.01~$\cal{M}_{\mathrm{J}}$, but perhaps
as large as 0.1~$\cal{M}_{\mathrm{J}}$, if the interaction with the stellar
magnetic field is important.  In turn, this limit implies a magnetic
field strength of only 10\% that of \objectname{Jupiter}.  
\cite{bs10} argue that Ohmic dissipation could explain the inflated
atmospheres of ``hot Jupiters,'' and they are able to explain the
atmospheres of \objectname{HD~209458b}, \objectname{HD~189733b}, and
\objectname{Tres-4b} assuming a field strength of~10~G at the cloud
tops, as compared to the value of~4.2~G for \objectname{Jupiter}.
Strictly, their estimate ($\sim 10$~G) is an upper limit because Ohmic dissipation need
not be the only effect operating to inflate ``hot Jupiter''
atmospheres.  
\cite{lhv10} show that the asymmetric transit properties of
\objectname{WASP-12b} might be able to be explained by a magnetopause,
if the planetary magnetic field strength is of order 1~\hbox{G}, but
their model also has a fairly weak dependence on the magnetic field
strength ($\propto B^{1/3}$).
Conversely, a small number of stars with ``hot Jupiters'' show
evidence for star-planet interactions \citep[e.g.,][]{sbwcc08}, which
appear to have a magnetic component.  
In contrast to ``hot Jupiters,'' with its large eccentricity orbit, \planet\ may be
immune to some of the effects that might act to damp the magnetic
field strength of ``hot Jupiters.''

The planet orbiting \hdstar\ passed  through periastron
on~HJD~2454424.86 (2007 November~20 at~08:05~\hbox{UT}).  This paper reports observations
at~325 and~1425~MHz ($\lambda$92~cm and $\lambda$21~cm, respectively) of
\planet\ close to its periastron.  In Section~\ref{sec:observe} we describe
the observations, analyses, and present the images obtained; in
Section~\ref{sec:discuss} we discuss our upper limits in the context of
recent multi-wavelength observations of \planet, and in
Section~\ref{sec:future} we make predictions about future radio observations.

\section{Observations}\label{sec:observe}

\begin{deluxetable}{lcccc}
\tablecaption{Observation Log\label{tab:log}}
\tablewidth{0pc}
\tablehead{%
 \colhead{Frequency} & \colhead{Wavelength} & \colhead{Bandwidth} & \colhead{Epoch}
	& \colhead{Orbital Phase}}
\startdata
 325~MHz & 90~cm & 12.5~MHz & HJD 2454423.89--2454424.22 & 0.69--0.73 \\
         &       &          & (2007 November~19 09:20--17:17 UT) \\
1425~MHz & 20~cm & 50~MHz   & HJD 2454424.74--2454425.07 & 0.94--0.12 \\
         &       &          & (2007 November~20 05:46--13:44 UT) & \\
\enddata
\end{deluxetable}

\hdstar\ was observed\footnote{
program AL715.
} at~325 and~1425~MHz with the Very Large Array (VLA), in its B
configuration, near the periastron passage
of the planet.  Table~\ref{tab:log} summarizes relevant details of the
observations, with specific information about the actual times of
observations and the planetary phase.  In addition to \hdstar, at both
frequencies, \objectname{3C~286} was observed to serve as a flux
density and bandpass calibrator, and the VLA calibrator
\objectname{J0929$+$502} was observed as a visibility phase
calibrator.  Although the VLA could operate at~74~MHz \citep{k+07},
the 74~MHz dipoles were not mounted at the time of these observations.

Standard data reduction techniques were used within the AIPS\footnote{
version 31DEC08.}
software package to calibrate the spectral bandpass, the visibility
amplitudes, and visibility phases.  The data were acquired in a
spectral-line mode both to enable the identification and excision of
radio frequency interference (RFI) and for the purposes of wide-field
imaging.  These issues of RFI excision and wide-field imaging are
particularly relevant for the 325~MHz observations.

\begin{figure}[tb]
\includegraphics[angle=-90,width=0.95\columnwidth]{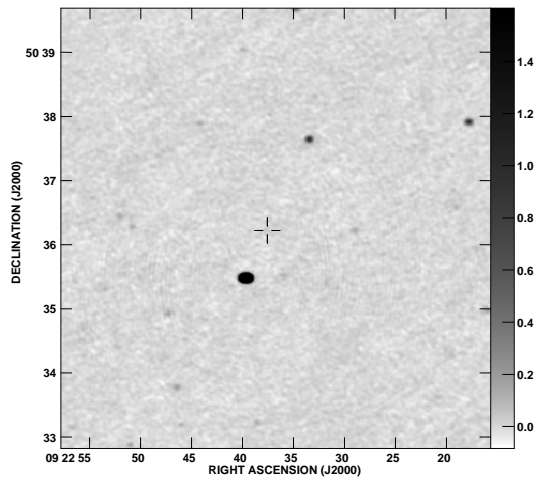}
\vspace*{-1ex}
\caption{Field around \hdstar\ at~1425~MHz on~HJD~2454425, which
  includes the time of periastron (Table~\ref{tab:log}).  The cross marks the location of the
star.  The beam for this image is 5\farcs49 $\times$ 4\farcs35, the
noise level in the image is 16~\ujybm, and the gray scale is
logarithmic between~$-80\,\ujybm$ and~1.6~\mjybm.  The source to the
southeast of the star is \protect\objectname{FIRST~J092239.6$+$503529}.}
\label{fig:20cm}
\end{figure}

\begin{figure}[bt]
\includegraphics[angle=-90,width=0.95\columnwidth]{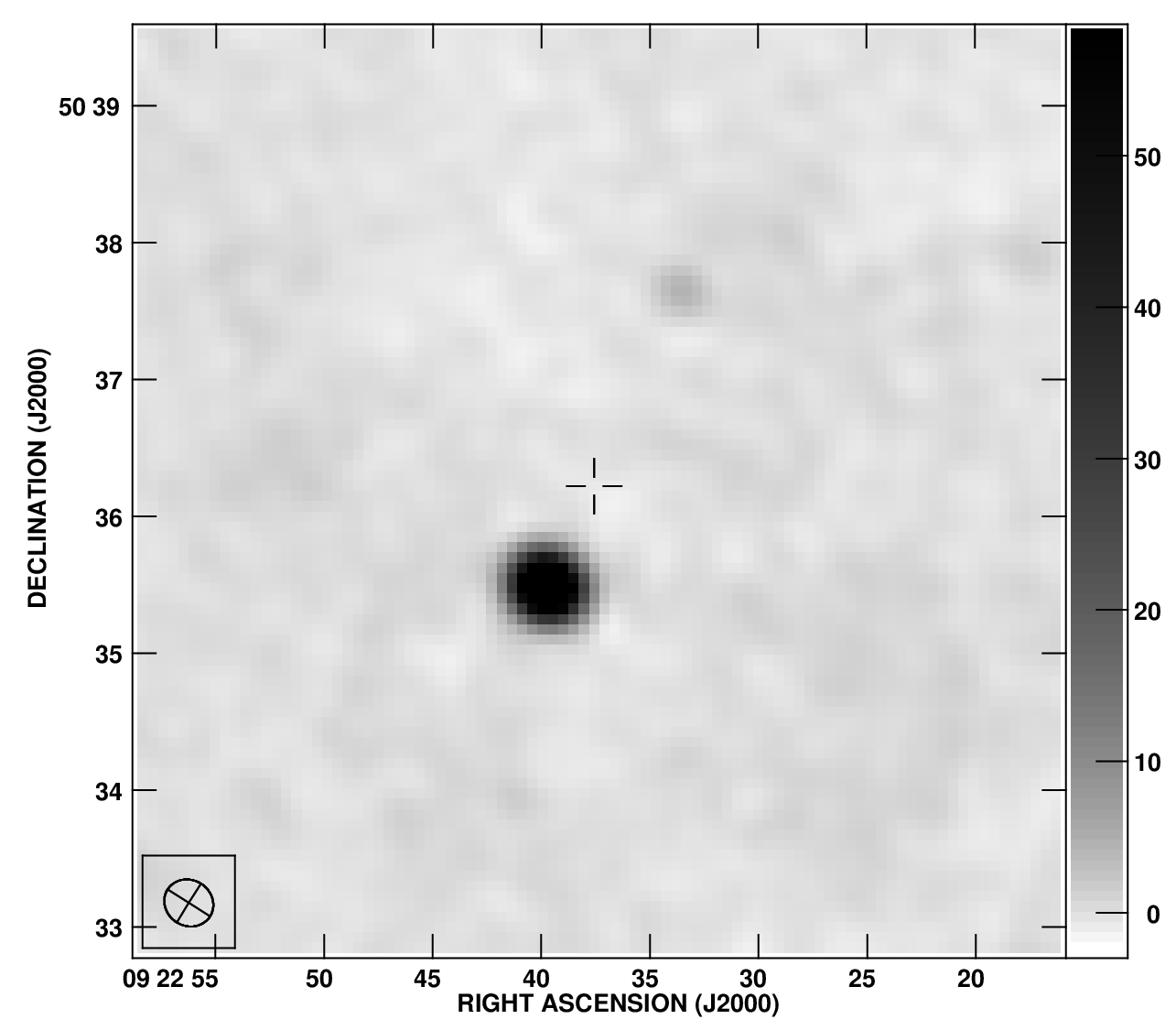}
\vspace*{-1ex}
\caption{Field around \hdstar\ at~325~MHz on~HJD~2454424, just before
the planet's periastron passage (Table~\ref{tab:log}).  The cross marks the location of the
star.  The beam for this image is 22\farcs1 $\times$ 20\farcs1 and is
shown in the lower left corner; the
noise level in the image is 580~\ujybm; and the gray scale is
logarithmic between~$-2.3\,\mjybm$ and~58~\mjybm.  The area shown is
approximately the same size as that shown in Figure~\ref{fig:20cm}.}
\label{fig:90cm}
\end{figure}

Following the data calibration, several iterations of imaging and
self-calibration were used to produce images of the field around
\hdstar; at both frequencies, polyhedral imaging was used to image the
primary beam field of view \citep{cp92}.  Figures~\ref{fig:20cm}
and~\ref{fig:90cm} present the inner portions of the resulting images,
centered on \hdstar.  The resulting image
noise levels are 16~\ujybm\ at~1425~MHz and 580~\ujybm\ at~325~MHz,
which in both cases are within approximately 50\% of the expected
thermal noise levels.

\section{Discussion}\label{sec:discuss}

At neither frequency do we detect statistically significant emission
from the direction of the star.  Table~\ref{tab:limits} summarizes the
resulting flux density limits on this system near periastron passage
and the resulting luminosity limits.  For reference, the nominal
luminosity of \objectname{Jupiter} is $2 \times
10^{18}$~erg~s${}^{-1}$ at a frequency near~40~MHz.

\begin{deluxetable}{lcc}
\tablecaption{Luminosity Limits for \planet\label{tab:limits}}
\tablewidth{0pc}
\tablehead{%
 \colhead{Wavelength} & \colhead{Flux Density Limit}
	& \colhead{Luminosity Limit} \\
	              & \colhead{(3$\sigma$)}
	& \colhead{(erg~s${}^{-1}$)} }
\startdata
1425~MHz & 48~$\mu$Jy & $2.7 \times 10^{23}$ \\
 325~MHz & 1.7~mJy    & $2.3 \times 10^{24}$ \\
\enddata
\tablecomments{In converting from flux density to luminosity, an
emission bandwidth of order the observing frequency is assumed.}
\end{deluxetable}

The age of \hdstar\ is uncertain, with estimates ranging from as
little as 0.3~Gyr to in excess of 14~Gyr
\citep{wmbv04,sgc05,vf05,tfsrfv07,hna09}, but with many estimates
favoring an age of approximately 7~Gyr.  For the purposes of this
discussion, we take the age of \hdstar\ to be comparable to that of
the \objectname{Sun}, with a comparable stellar wind strength.
Consequently, we assume that a Jovian planet, with an orbital
semi-major axis comparable to that of \objectname{Jupiter} ($\approx 5$~AU), would
have a radio luminosity similar to that of \objectname{Jupiter}.

It is clear that our luminosity limits remain several orders of
magnitude ($10^5$--$10^6$) above the luminosity of \objectname{Jupiter}.  The
original motivation for these observations was the expectation that
the high eccentricity of the planet's orbit could produce a dramatic
variation in the strength of the planet's radio emission.  Because a
planet's magnetospheric emission is controlled by the input stellar wind
power, the luminosity of the planet is a strong function of its
distance from the star.  \citet[Appendix~A]{fdz99} discuss the scaling
of the luminosity with distance and find that an approximate scaling
law is $d^x$, with $x \approx -1.6$.  
\cite{gzs07,gpkmmr07} consider more complex stellar wind
models, but a strong dependence on the planet-star distance remains.

Over the course of the planet's orbit, from apastron to periastron, we
therefore expect that the luminosity of \planet\ could vary by a
factor of~200.  Even at apastron (0.87~AU), \planet\ is still 
closer to \hdstar\ than \objectname{Jupiter} is to the
\objectname{Sun}.  If we consider the further boost that could be
obtained by moving a Jovian-like planet from a distance of~5~AU
to~0.87~\hbox{AU}, the boost would be a factor of~16.  Thus, one might
expect that, near periastron, \planet\ could be as much as a factor of
3000 times more luminous than \objectname{Jupiter}.  However, even an
increase by a factor of order 3000 over Jupiter's luminosity would be
insufficient to increase \planet\ to a level comparable to our
detection thresholds.

A simple explanation for our non-detections is therefore 
that, even with the added boost due to the small planet-star distance
during periastron, the planet's luminosity remains too low to be
detected with current equipment.  There are several other factors,
however, that could also explain our non-detection.

One possibility is that the radio emission is not beamed in our
direction.  In the case of \objectname{Jupiter}, \cite{zck04}
determined that it emits into a solid angle of approximately 1.6~sr,
directed roughly along Jupiter's rotational axis.  \planet\ is one of
the rapidly growing class of transiting extrasolar planets, and recent
transit observations demonstrate that there is a significant
spin-orbit misalignment ($\sim 50\arcdeg$) in this system
\citep{fwk09,g-mm09,mhb+09,phi+09,whj+09}.  For a transiting planet
whose rotational axis is aligned with the stellar rotation axis and
\emph{no} spin-orbit misalignment, the probability of intercepting the
emission beam would be relatively small.  However, the existing
spin-orbit misalignment in this system increases the probability that
emission beam intersects the line of sight.  Of course, it is possible
that the emission cone of \planet\ is much smaller than that of
\objectname{Jupiter}, such that the probability of intercept remains
quite low.

A second possibility is that the expected upper frequency limit of
emission is (well) below our observation frequencies.  The intense
cyclotron maser emission of \objectname{Jupiter} cuts off dramatically at
frequencies above approximately 40~MHz, as the local plasma frequency
in the polar emitting region exceeds the electron cyclotron frequency
and the radio emission cannot escape.  For most extrasolar planets,
one can estimate the upper frequency limit by scaling their magnetic
moments from that of \objectname{Jupiter}.  Previous estimates for the
magnetic moment, and upper frequency limit, for \planet\ have varied
wildly.  \cite{lf+04} estimated an upper frequency limit of around
180~MHz, while \cite{gzs07} estimated 0.8~MHz.

The combination of the eccentric orbit and the planetary transits,
however, now provide sufficient information to estimate, with relatively
low uncertainty, the upper emission frequency.  This upper emission
frequency is given by \citep{fdz99,lf+04}
\begin{equation}
\nu_{\mathrm{c}}
 \approx 24\,\mathrm{MHz}\left(\frac{\omega}{\omega_{\mathrm{J}}}\right)\left(\frac{M}{M_{\mathrm{J}}}\right)^{5/3}\left(\frac{R}{R_{\mathrm{J}}}\right)^3
\label{eqn:freq}
\end{equation}
for a given planetary mass~$M$, rotation and radius~$R$.  In practice,
this value may underestimate the actual upper emission frequency, as
the case for \objectname{Jupiter} is $\nu_{\mathrm{c}} \approx 40$~MHz
whereas Equation~(\ref{eqn:freq}) suggests 24~MHz.  This expression
also assumes that the planetary magnetic dipole is not offset
significantly (as it is for the planets \objectname{Uranus} and
\objectname{Neptune}), which can produce significant asymmetric emission.

Of the various determinations of the mass and radius of the planet
\citep{fwk09,g-mm09,mhb+09,phi+09,whj+09}, we adopt those that
\cite{phi+09} obtained from a Bayesian analysis of all of the data
available to them.  They found a planetary mass of $3.94 \pm
0.11$~$M_{\mathrm{J}}$ and a radius of $0.98 \pm
0.03$~$R_{\mathrm{J}}$, implying a density about four times higher than
that of \objectname{Jupiter}; their models also suggest that the core
mass of \planet\ could be several times that of \objectname{Jupiter}.

\cite{h81} discusses the tidal effects in eccentric systems.  The
large eccentricity for the orbit of \planet\ should have driven it
into a state of pseudo-synchronization between its rotation and
orbital periods.  Using expressions derived by \cite{h81}, we estimate
a (pseudo-synchronized) rotational period of~39.9~hr.

Evaluating Equation~(\ref{eqn:freq}), we find that the upper emission
frequency of approximately 55~MHz, and potentially as high as 90~MHz,
if one accounts for the 50\% difference between the actual value for
\objectname{Jupiter} and that predicted by Equation~(\ref{eqn:freq}).
It is clear that this upper emission frequency is well below that of
our observations.

This estimate for the emission frequency differs from that of
\cite{lf+04} by only a factor of~2--3, but it differs from that
obtained by \cite{gzs07} by a factor of order $10^2$.  We believe that
the differences can be attributed to assumptions of the rotation
period.  \cite{lf+04} assumed a rotation period equal to that of
\objectname{Jupiter}, which is within a factor of~4 of the rotation
period we adopt here.  However, \cite{gzs07} identified the planet as
``tidally locked,'' which would mean that it would have an extremely
slow rotation rate, and therefore a low magnetic moment and low
emission frequency.  One of the key factors that may have led
\cite{gzs07} to identify the planet as tidally locked is that they
appear to have used expressions for estimating the synchronization
time scale which assume a circular orbit (their Appendix~B).

\section{Future Prospects}\label{sec:future}

We consider the possibilities of future observations from two
perspectives.  First, are there planets other than \planet\ that
should be targeted? Second, what are the prospects for future
instruments to detect \planet?

While, \planet\ is not the only known planet with a high-eccentricity
orbit, it appears the most promising for future radio observations.
\objectname{HD~20782b} is a 1.9~$M_{\mathrm{J}}$ (minimum mass) planet
in an orbit with an eccentricity of~0.97 and an orbital period
of~592~days \citep{otjbmcb09}, and \objectname{HD~4113b} is a
1.6~$M_{\mathrm{J}}$ (minimum mass) planet in an orbit with an
eccentricity of 0.903 and an orbital period of 527~days
\citep{tsu+08}.  The pseudo-synchronization rotational period is
coupled to the orbital period, so these longer orbital periods imply
substantially larger rotational periods, as compared to \planet.  We
estimate rotational periods of 63.8~hr and 14~days for
\objectname{HD~20782b} and \objectname{HD~4113b}, respectively.
Combined with their lower masses, Equation~(\ref{eqn:freq}) suggests
that their emission frequencies are likely to be below~20~MHz.  Unless
future surveys reveal planets with masses and orbital eccentricities
comparable to those of \planet\ \citep[but see][]{otjbmcb09}, it
appears likely to remain the most promising target for radio
observations of (high-eccentricity) planets.

With respect to observations of \planet, our analysis suggests that
future observations must be both more sensitive and at a lower
frequency than those that we report here.  There are several observing
systems that are either under construction or in the design phase
that we now consider.

The Expanded Very Large Array (EVLA) project exposed certain
shortcomings of the VLA's 74 MHz
system \citep{k+07}, primarily its generation of harmonics at higher frequencies
that fall within the EVLA's operational spectral window.
Consequently, the NRL has funded the National Radio Astronomy
Observatory to develop a new suite of low-frequency receivers.  With
the previous 74 MHz VLA system, automated data processing of
approximately a 1 hr observation could obtain rms intensity levels of
order 100~\mjybm\ \citep{clc+07}, while more focused efforts on
longer observations could obtain noise levels of~25~\mjybm\
\citep{crjkl04}.  At a minimum, the new ``EVLA-low band'' system will
have a substantially larger bandwidth.  There has also been
continuing algorithmic improvements in the areas of RFI
identification and excision and ionospheric calibration
\citep[e.g.,][]{a09,ivdtccvbr09,odbbzbp10}.  
As an ansatz, we assume that a
combination of algorithmic improvements and the larger bandwidth could produce an
approximate factor of~10 improvement in sensitivity, i.e., 3~\mjybm.
The resulting (3$\sigma$) luminosity limit would then be
$10^{23}$~erg~s${}^{-1}$ (for an assumed 10~MHz intrinsic emission
bandwidth of the electron cyclotron maser emission).  This value is
comparable to our more stringent limit (at~1425~MHz), but it would be
at a frequency at which emission might more likely be detected.

The Low Frequency Array
({LOFAR}\footnote{%
http://www.astron.nl/radio-observatory/astronomers/lofar-astronomers}
)
is currently under construction in the Netherlands and other European
countries.  When complete, it will consist of many dipole phased-array
stations, operating in a ``high band'' (HBA) and a ``low band'' (LBA).  LOFAR-LBA
band will cover the frequency range 30--80~MHz, with a peak sensitivity around~60~MHz,
well matched to the expected emission frequency of \planet.  Current
estimates for the sensitivity of LOFAR-LBA are that it will
produce a similar or slightly better thermal noise level
(about~7~\mjybm\ in a 1 hr integration) as the EVLA-low band system.
We emphasize that the limiting factor in imaging sensitivity might
very well not be the thermal noise level, but other factors such as
ionospheric calibration.  Nonetheless, it appears that the LOFAR-LBA
band system would produce similar constraints.

Following the EVLA-low band, \hbox{LOFAR}, and other pathfinder
systems are likely to be the low-frequency component of the Square
Kilometre Array (SKA-lo) and ultimately a low frequency array on the
Moon (the Lunar Radio Array).  The current design goal for SKA-lo has
a lower operational frequency of~70~MHz with enough collecting area
that the notional thermal noise level would be of order 10~\ujybm\ in
a 1 hr integration.  Unfortunately, candidate sites for the SKA-lo
are in the southern hemisphere (Australia and South Africa,
terrestrial latitude $\approx -30\arcdeg$), from which \planet\ cannot
be observed.  If future planetary surveys detect other \planet-like
planets and if the thermal noise level can be approached with the
SKA-lo, the implied luminosity limit would then be $3 \times
10^{20}$~erg~s${}^{-1}$, well within the range we estimate might be
plausible based on the larger stellar wind loading of {\planet}'s
magnetosphere.

\acknowledgements

We thank G.~Laughlin for the initial inspiration for these
observations and for pointing out that \planet\ should be in a state
of pseudo-synchronization; J.~Schneider for the Extrasolar Planets
Encyclopedia; and P.~Perley and W.~Koski for helpful discussions.
T.{}J.{}W.{}L.\ thanks \hbox{USNO}, where much of this paper was written, for its
hospitality.
The National Radio Astronomy Observatory is a facility of the National
Science Foundation operated under cooperative agreement by Associated
Universities, Inc.
The {LUNAR consortium} is funded by the NASA Lunar Science
  Institute (via Cooperative Agreement NNA09DB30A) to investigate
  concepts for astrophysical observatories on the Moon.
Basic research at NRL is supported by 6.1 Base funding.

\end{document}